\documentclass{wow3}
\usepackage{amsmath}
\usepackage{amssymb}
\usepackage{graphicx}
\usepackage{mathrsfs}

\def\Real{\mbox{Re}}      % cf plain TeX's \Re and Reynolds number
\def\Imag{\mbox{Im}}      % cf plain TeX's \Im
\newcommand{\rij}[2]{\mbox{$\overline{u'_{#1}u'_{#2}}$}}

\newcommand{\beqa}{\begin{equation}}
\newcommand{\eeqa}{\end{equation}}
\newcommand{\no}{\nonumber}
\newcommand{\be}{\begin{eqnarray}}
\newcommand{\en}{\end{eqnarray}}

\newcommand{\df}[2]{\displaystyle\frac{#1}{#2}}
\newcommand{\tf}[2]{\textstyle\frac{#1}{#2}}

\newcommand{\PDD}[2]{\df{\partial #1}{\partial #2}}

\newcommand{\eps}{\varepsilon}

\newcommand{\Half}{\mbox{\tiny $\tf{1}{2}$}}

\title[Growth of unsteady wave groups by shear flows]{Growth of unsteady wave groups by shear flows}
\author{S.G. Sajjadi$^{1,2}$, J.C.R. Hunt$^{2,3}$ and F. Drullion$^{1}$}
\affiliation{$^{1}$\, Department of Mathematics, ERAU, Florida, USA.\\
$^{2}$\, Trinity College, University of Cambridge, UK.\\
$^{3}$\, CPOM, Department of Earth Sciences, UCL, UK.}

\begin{document}

\pagenumbering{arabic}
\setcounter{page}{79}

\maketitle

\begin{abstract}
A weakly nonlinear theory has been proposed and developed for calculating the energy-transfer rate to individual waves in a group. It is shown what portion of total energy-transfer rate, over the envelope of wave group, affects individual waves in the group. From this an expression for complex phase speed of individual waves is calculated. It is deduced that each wave in a group does not grow at the same rate.
It is shown that the critical layer is no longer symmetrical compared with the ideal
monochromatic waves. This asymmetry causes the critical layer height to be lower
over the downwind part. Therefore the positive growth of the individual waves on the
upwind part of the wave group exceeds the negative growth on the downwind part
(which would not be true if $z_c$, where the mean flow $U$ is equal to the speed of the wave propagation, was the same over the whole group). This leads to the
critical layer group effect producing a net horizontal force on the waves, in
addition to the sheltering effect. Computational simulations over a non-growing wave group is also
presented, which confirms the above postulation made by Sajjadi, Hunt and Drullion (2014) (SHD).
\end{abstract}

\vskip0.3in

\section{Introduction}
The determination of the total energy-transfer rate, $\beta_{g}$, from wind shear flow to a group of waves is of great interest and importance.  Since a group of waves consists of various waves of different amplitude and steepness with its envelope.  It is important to calculate what proportion of $\beta_{g}$ affect individual waves present in the group.  Only in this way can the growth of a wave group can properly be calculated

In general the growth rate of a wave, $\zeta$, is proportional to the energy-transfer rate which in turn depends upon the imaginary part of the wave complex phase speed, $c_i$.  Thus, $\zeta=kc_i$, where $k$ is the wave-number.  Hence, if there are $N$ waves which constitute a group then each wave grows at the rate $\zeta_\ell$ for $\ell=1,2,...,N$  This means that the energy-transfer rate to each wave is $\beta_\ell$ portion of $\beta_{g}$. 

In this paper, we present some preliminary results of an on going investigation of shear flow of wind blowing over wave groups.  We first outline our theory for determining $\beta_\ell$ and then show results of numerical calculations of turbulent shear flow over a wave group.  For the latter, the mean flow features and turbulence statistics of cat's-eye and separated wake region are computed using a nonlinear fully realizable Reynolds stress closure model which includes the effects of inhomeogenous, anistotropic distortion, diffusion, and dissipation. 

\section{Growth of waves within a group}
When the side-band amplitudes of progressive waves, of finite amplitude on deep water (namely, Stokes wave) become nearly equal, as they always do after sufficient amplification (Benjamin-Feir instability) profile of the forward-traveling modulation of the primary wave train may be expressed as 
\be
\varsigma=\Real\left\{ae^{ik(x-ct)}+\varepsilon(t)\left[e^{ik_2(x-c_2t-\theta_2)}+e^{ik_3(x-c_3t-\theta_3)}\right]\right\}
\equiv\eta+\varepsilon(t)(\eta_2+\eta_2)\label{1}
\en
where 
$$\varepsilon(t)=\exp\left\{\tf{1}{2}\delta(2k^2a^2-\delta^2)^{\Half}\omega t\right\},\quad k_{2,3}=k(\kappa\pm 1),\quad\omega_{2,3}=\omega(\delta\pm 1),$$ 
with both $\kappa$ and $\delta\ll 1$.  As was pointed out by Benjamin and Feir (1967), the terms in $\varepsilon(t)$ represent a gradual modulation, at wave-number $\kappa k\ll k$ of the fundamental wave $a \cos(kx-\omega t)$.  Note that, the first term, which is proportional to $\cos(\tf{1}{2}\theta_2)$, describes an amplitude modulation, and the second term with rapidly varying factor, describes a phase modulation.  Then the modulation envelope propagates at the group velocity $c_g=\delta\omega/\kappa k=\tf{1}{2} c$, where $c=c_r+ic_i$.

\begin{figure}
   \begin{center}
\includegraphics[width=12cm]{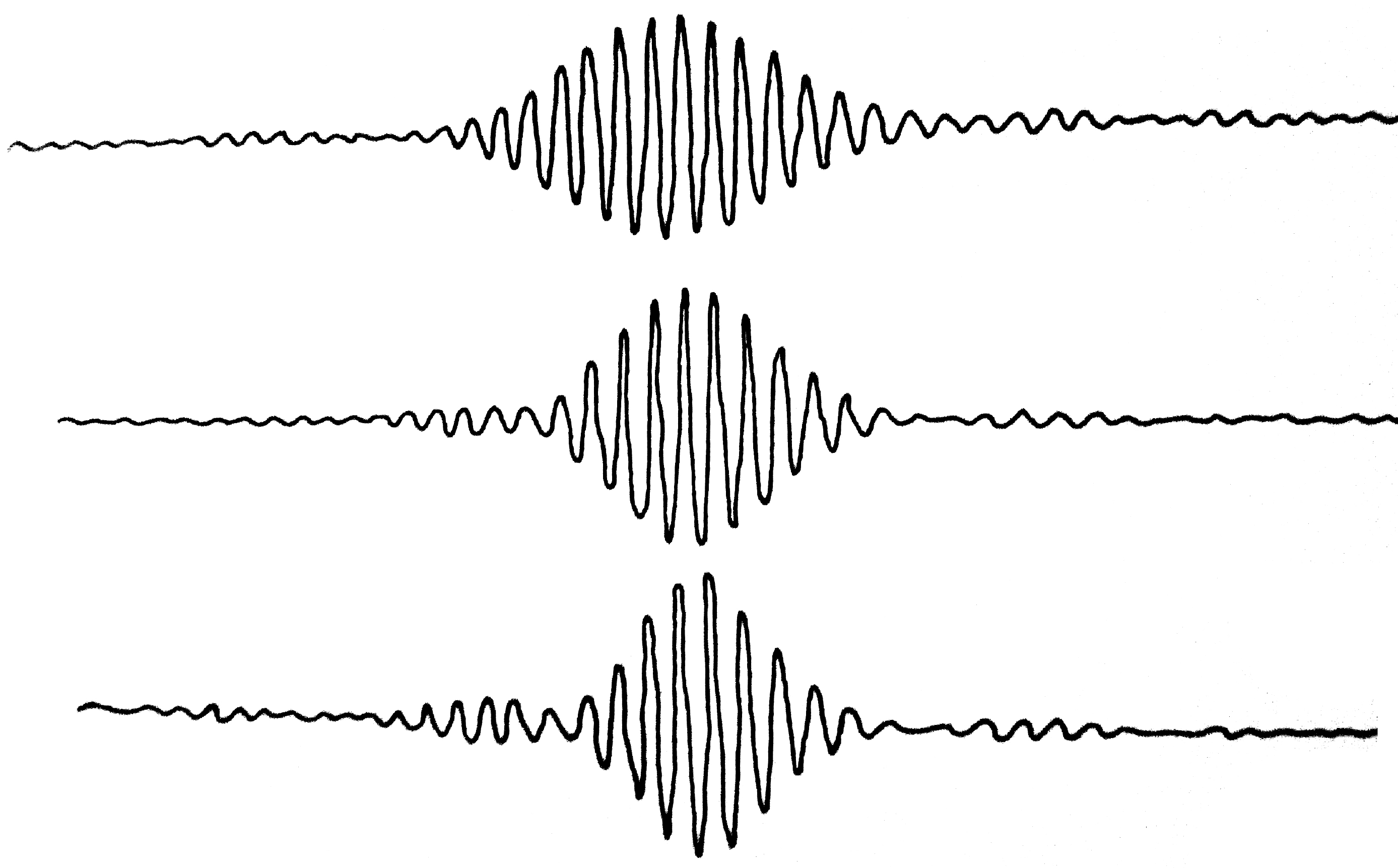}
   \end{center}
\caption{\footnotesize Schematic diagram of wave groups for different values of $\kappa$ and $\delta$ which may be represented by equ. (\ref{1}). Taken from Feir (1967).}
\end{figure}

Following Miles (1957), we shall assume an inviscid flow above the wave group (1) whose mean velocity profile is logarithmic.  Then, the surface pressure may be posed in the form (c.f. Miles 1957, equ. (2.3))
\be 
p_a=\rho_aU^2_1k\left\{(\alpha+i\beta)\eta+\varepsilon(t)\left[(\alpha_2+i\beta_2)\eta_2+
(\alpha_3+i\beta_3)\eta_3\right]\right\}\label{2}
\en 
then the complex phase speeds can be easily be shown to be
\be 
c_{2,3}=c_{2w,3w}\left[1+\df{1}{2}\df{k^2}{k^2_{2,3}}(\alpha_{2,3}+i\beta_{2,3})\left(\df{U_1}{c_{2w,3w}} \right)^2s\right]\label{3}
\en 
where $s\equiv \rho_a/\rho_w\ll1$, the suffices $a$ and $w$ refer to the air and water, respectively, and $U_1=U_*/K$.  Here, $U_*$ represents the wind friction velocity and $K$ is the von-K\'arm\'an's constant. 

In equation (\ref{2}), $\alpha_\ell$ and $\beta_\ell$, for $\ell=1,2,3$ are the real and imaginary parts of the interfacial impedance for each wave $l$ in a group.  Also
\be 
c_{2w,3w}=[g(1+k^2a^2)/k]^{\Half}-2ik_{2,3}\nu_w;\quad|k_{2,3}\nu_w/c_{2,3}|\ll 1\label{4}
\en 
are wave speeds in the absence of the air, and $\nu_w$ is the kinematic viscosity of the water. 

The aim here is to determine the total energy-transfer rate, $\beta_g$, over a wave group and thence deduce the contribution that $\beta_g$ makes to energy-transfer rate $\beta_l$ for each $i$ number of waves in the group.  Thus, for a growing wave group this requires determination of the imaginary part of waves phase velocity in a group.  We remark that the total energy-transfer rate is related to the complex amplitude of the normal, ${\mathscr P}_g$ and the shear ${\mathscr T}_g$, stresses through
\be 
\alpha_g+i\beta_g\equiv \df{c^2_g-c^2_{gw}}{sU^*_1}=\df{({\mathscr P}_g+i{\mathscr T}_g)_0}{kaU^2_1}\label{5}
\en 
where the suffix zero implies evaluation on the surface of a wave group. 

This task requires extending the resonant interaction theory of Longuet-Higgins (1962), for two gravity waves, to several waves, in conjunction with phase velocity effects intertiary wave interaction (Longuet-Higgins and Phillips 1962).  Thus, in the present case, applying the above theories we find that
\be 
{\mathscr C}_g=\tf{1}{2}g\left[2\sqrt{gk(1+\kappa)}-\sqrt{gk(1-\kappa)}\right]^{-1}\label{6}
\en 
for the two side-band waves.  Then, for example, the interaction of the lower side-band wave with the primary wave yields
\be 
C_g=\tf{1}{2}g\left[2\sqrt{gk(1+k^2a^2)}-\sqrt{gk_3}\right]^{-1}\label{7}
\en 
Once all interactions between various waves are taken into account, it is relatively easy (though algebraically lengthy) to show the end result of summing of all the sub-group velocities, given by (\ref{5}) and (\ref{6}), yields {\it exactly} $c_g=\tf{1}{2}c$.  However, it must be noted that as $c$ is complex then so will be $c_g$, and the imaginary part $c_{gi}\equiv\Imag(c_g)$ will be responsible for the growth of the group envelope. We remark, however, that according to above expressions not all waves in a group grow at the same rate.  

\vskip0.3in
\section{Determination of $c_{gi}$}

As in SHD, we model the perturbations to airflow by eddy viscosity $\nu_e$.  The amplitude of the vertical velocity perturbation satisfies the inhomogeneous Rayleigh equation
\be 
\PDD{^2\hat{\mathscr W}_\ell}{z^2}-\left(k^2+\df{U''}{U-ic_i}\right)\hat{\mathscr W}_\ell=\df{i}{U-ic_i}\PDD{^2}{z^2}\left(\nu_e\PDD{^2\hat{\mathscr W}_\ell}{z^2}\right)\label{8}
\en 
As was originally suggested by Belcher, Hunt and Cohen, in the middle layer where the advection is much smaller then the curvature term, equation (\ref{8}) reduces to 
\be 
\PDD{^2\hat{\mathscr W}_\ell}{z^2}-\df{U''}{U-ic_i}\hat{\mathscr W}_\ell\sim 0\label{9}
\en 
and the solution is regular since $U>0$.  We note that if $c_i=0$, then significantly is resolved by inertial effects.

Next, following Beji and Nadaska (2004) the amplitude of the surface pressure, for the present case, may be written as
\be 
& &[P_1+k\varepsilon(P_2+P_3)]_a=[P_1+k\eps(P_2+P_3)]_0-2\rho_a ga (2-k\varepsilon)+
\no\\
& &\rho_a kac^2\left(\df{k}{c}\right)\left\{\left[\df{1}{\hat{\mathscr W}_0}\int^{\infty}_{z_0}(U-ic_i)\hat{\mathscr W}\,dz\right]
+2k\eps\left[\sum^{3}_{\ell=2}\df{1}{(\hat{\mathscr W}_\ell)_0}\int^{\infty}_{z_0}(U-ic_i)\hat{\mathscr W}_\ell\,dz\right]\right\}\no\\\label{10}
\en 
where now the suffix zero indicates evaluation at the surface roughness $z=z_0$ and the first term on the right-hand side of (\ref{10}) is the total atmospheric pressure at the surface.  Equation (9) represents the air pressure on the surface of the wave group due to the wind. 

The continuity of pressure across the air-sea interface yields to the following expressions 
\be 
& &[P_1+k\varepsilon(P_2+P_3)]_0-\tf{1}{2}\rho_w ga(2+k\eps)+\rho_w kac^2(1-2k\varepsilon)=
\no \\
& &[P_1+k\eps(P_2+P_3)]_0-\tf{1}{2}\rho_a ga(2+k\varepsilon)+\rho_a kac^2[T_1+2(T_2+T_3)]\no
\en 
where $T_1=\df{k/c}{\hat{\mathscr W}_0}\int^{\infty}_{z}(U-ic_i)\hat{\mathscr W} dz$ with similar expressions for $T_2$ and $T_3$.  Eliminating $[P_1+k\varepsilon(P_2+P_3)]_0$, dividing by $\rho_w a$, and solving for $c^2$ we obtain
$$c^2=c_r\{(1+\varepsilon)+\tf{1}{2}sI_c-\tf{1}{8}\varepsilon(2+s-sI_c)\}+O(\varepsilon^2,s^2)$$ where $I_c=T_1+2(T_2+T_3)$ and $c_r=\sqrt{g(1+k^2a^2)/k}$.

Once $\hat{\mathscr W}_\ell$ is determined (as in SHD) then the complex integrals $I_c$ and hence the complex phase speeds (in conjunction with expressions (\ref{2})) can be evaluated. 

The total energy-transfer rate $\beta_g$ can be calculated from the following expression:
\be 
\beta_g=(c_r/U_1)^2\Imag\{I_c\}\label{11}
\en 
Finally, upon expansion of $\beta_g$ to $\beta_\ell$, for $\ell=1,2,3$, etc. the energy-transfer rate to each wave in a group may be determined. 

\section{Computations of turbulent flow over a wave group}

Typically in the ocean waves move in groups, which affects how
the wind flows over the waves, how the waves break and thence how droplets form.
This weakly nonlinear interaction of mechanisms significantly influence the average
momentum, heat and mass transfer associated with waves.

By considering the dynamics of typical wave groups, it becomes
possible to estimate rationally how air flow affects the nonlinear interactions between
waves, and compare how this relates to the wave-wave hydrodynamic interactions,
that are assumed to dominate the distribution of ocean waves. Thus variations of wave
shapes within a group could also affect the net wave growth, while violent erratic winds
can prevent the formation of wave groups, so that wave growth may be reduced (but
spray from waves is increased) as is observed near the center of hurricanes.
At higher wave speeds, another mechanism is also significant, namely the displacement 
of the critical layer outside the surface shear layer (i.e. $c_r > U_*$).

The main objective of the present computations is to develop a computational model
for turbulent shear flow of air over steadily moving wave groups in which individual
waves are unsteady. Firstly this shows, by linear theory, for the individual waves in the
group the combined effects of the unsteady critical layer flow, and the viscous/turbulent
sheltering on the lee sides of the wave. Secondly, by using weakly nonlinear theory to
analyze the disturbed air flow over the waves in groups, it is shown how the air speed
over the downwind part of the group is lower over the upwind part. This asymmetry
causes the critical layer height to be lower over the downwind part, which leads to
critical layer effect producing a net horizontal force on the waves, in addition to the
sheltering effect.

\begin{figure}
   \begin{center}
\includegraphics[width=12cm]{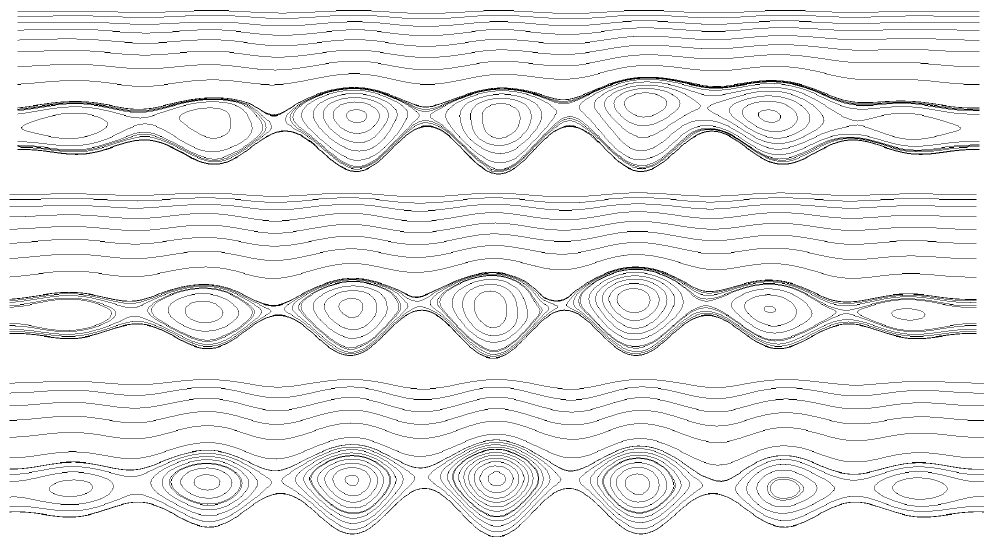}
   \end{center}
\caption{\footnotesize Turbulent flow over a group of wave for $c_{gr}/U_*=3.9$ (the top); 
7.8 (the middle); and 11.5 (the bottom).}
\end{figure}

The concepts based on this analysis will indicate how the effects of wave dynamics
will also affect the wave growth. On the other hand we shall investigate whether violent
erratic winds prevent the formation of wave groups, thereby reducing the wave growth.
A number of mechanisms have been proposed for how air-flow over a horizontal
body of liquid produces waves on its surface. Most of those proposed have been linear
and therefore can be applied to any spectrum of waves. However, the mechanisms and models
based on them are regularly applied when the surface disturbances induce gas and
liquid flows that are nonlinear, and where the waves are not monochromatic. 

When the waves begin to grow at a rate $kc_i$, comparable with $U_* k$, the critical layer is
outside the shear layer near the surface, and the dynamics across the critical layer are
determined by inertial forces as the flow accelerates and decelerates over the wave. We remark
only if the wave is growing (or decaying), i.e. $c_i \neq 0$, is there a net force on the wave
caused by critical layer dynamics (SHD). Their conclusion from
a triple deck analysis agrees with the result of Miles' (1957) different method of analysis,
but not his, nor many subsequent authors’ (e.g. Lighthill 1962), conclusion that 
there is a net inviscid force on monochromatic non growing waves, (i.e. $c_i=0$).

The way we propose to resolve this controversy is to consider weakly nonlinear interactions of
different mechanisms, such varying wave shape, wave surface, turbulence stresses, etc.
As wave groups are the best way of
looking at the nonlinear mechanisms, SHD analyzed the perturbed flow in the inertial
layers at the critical heights and showed how it contributes to the growth and decay
of waves over the upwind and downwind parts of a wave group. They concluded
(against Miles 1957, Lighthill 1962) that there would be no net growth, as a result of the
critical layer mechanism, if the airflow was linearly related to the surface. But a weakly
nonlinear air flow analysis can show how the airflow over the wave group becomes
asymmetric and affects to a small extent the heights of the critical layers over the group.

To demonstrate some of the arguments just made, we present some preliminary computations
of turbulent flow above a group of waves whose profile is calculated from equation (\ref{1}). 
We have adopted a fully realizable Reynolds-stress turbulence model 
\begin{eqnarray}
\frac{D \rij{i}{j}}{Dt} = P_{ij}+\Pi_{ij}-\varepsilon_{ij} +d_{ij}\label{12}
\end{eqnarray}
for computing $\rij{i}{j}$ which appear in the Reynolds-averaged Navier-Stokes equations.
The details of the closure model for various term in equation (\ref{12}) can be found in Sajjadi {\it et al.} (2001). 

In figure 2 we have shown preliminary results of our computations for three different values of
$c_{gr}/U_*$, where $c_{gr}$ is the real part of the group velocity $c_g$.  The values $c_{gr}/U_*$ used for computations depicted in figure 2 are 3.9 (the top); 7.8 (the middle); and 11.5 (the bottom). The results shown in figure 2 seems to confirm the points made in this section.

\section{Conclusions}

A conceptual theoretical and computational model has been developed  for the quasi-inviscid and turbulent shear flow over steadily moving wave groups, whose significance was first pointed out by M.E. McIntyre.
A weakly nonlinear theory is proposed to calculate the energy-transfer rate to individual waves in a group.
A computational model is used to analyze the disturbed turbulent air flow over the waves in
groups, which shows how the air flow over the downwind part of the group is lower
than over the upwind part. This asymmetry causes the critical layer height to be lower
over the downwind part. Therefore the positive growth of the individual waves on the
upwind part of the wave group exceeds the negative growth on the downwind part
(which would not be true if $z_c$ was the same over the whole group). This leads to the
critical layer group effect producing a net horizontal force on the waves, in
addition to the sheltering effect.


\begin{thebibliography}{}

\bibitem[Beji \& Nadaska (2004)]{BEJ} \textsc{Beji, S. \& Kadaoha, K.} 2004 {Solution of Rayleigh's instability equation for arbitrary wind profiles.} \textit{J. Fluid Mech.} \textbf{500}, 65--73.
 
\bibitem[beletal (1999)]{belet}
\textsc{Belcher, S.E.,  Hunt, J.C.R. \& Cohen, J.E.} 1999
{Turbulent flow over growing waves.} 
In {\it Proceedings of IMA Conference on Wind Over Waves}. Sajjadi, S.G.,  Thomas, N.H. \& Hunt, J.C.R. (eds). Oxford University Press, pp. 19--30. 
 
\bibitem[Benjamin \& Feir (1967)]{BF67} \textsc{Benjamin, T.B. \& Feir, J.E.} 1967 {The disintegration of wave trains on deep water. Part 1. Theory.} \textit{J. Fluid Mech.} \textbf{27}, 417--430.

\bibitem[Feir (1967)]{Feir67} \textsc{Feir, J.E.} 1967 {Discussion: some results from wave pulse experiments.} \textit{Rroc. Roy. Soc. Lond. A.} \textbf{299}, 54--58.
 
\bibitem[Lighthill (1962)]{Light} \textsc{Lighthill, M.J.} 1962 {Physical interpretation of
the theory of wind generated waves.} \textit{J. Fluid Mech.} \textbf{14}, 385--398. 

\bibitem[Longuet-Higgins (1962)]{MSLH} \textsc{Longuet-Higgins, M.S.} 1962 {Resonant interactions between two trains of gravity waves.} \textit{J. Fluid Mech.} \textbf{12}, 321--332.

\bibitem[Longuet-Higgins \& Phillips (1962)]{LHPhil} \textsc{Longuet-Higgins, M.S. \& Phillips, O.M.} 1962 {Phase velocity effects in tertiary wave interactions.} \textit{J. Fluid Mech.} \textbf{12}, 333--336.

\bibitem[Miles (1957)]{M57} \textsc{Miles, J.W.} 1957 {On the generation of surface waves by shear flows.} \textit{J. Fluid Mech.} \textbf{3}, 185--204.

\bibitem[Sajjadi et al (2001)]{sajetal} \textsc{Sajjadi,  S.G., Craft, T.J. \& Feng, Y.} 2001 {A numerical  study  of turbulent flow over a two-dimensional  hill.} \textit{Int.  J. Numer.  Meth.  Fl.} \textbf{35}, 1--23.

\bibitem[SHD (2014)]{SHD} \textsc{Sajjadi, S.G., Hunt, J.C.R. \& Drullion, F.} 2014 {Asymptotic multi-layer analysis of wind over unsteady  monochromatic surface waves.} \textit{J. Eng. Math.} \textbf{84}, 73--85.  

\end{thebibliography}
\end{document}